\documentclass[doublecol]{epl2} 

\title{General coevolution of topology and dynamics in networks}
\shorttitle{General coevolution of topology and dynamics in networks} 
\author{J. L. Herrera\inst{1} \and M. G. Cosenza\inst{2} \and K. Tucci\inst{2} \and J. C. Gonz\'alez-Avella\inst{3,4}}
\shortauthor{Herrera \etal}
\institute{                    
\inst{1} Departamento de C\'alculo, Escuela B\'asica de Ingenier\'ia, Universidad de Los Andes, M\'erida, Venezuela\\
\inst{2} Centro de F\'isica Fundamental, Universidad de Los Andes, M\'erida, Venezuela\\
\inst{3} IFISC, Instituto de F\'isica Interdisciplinar y Sistemas Complejos (CSIC-UIB), E-07122 Palma de Mallorca, Spain\\
\inst{4} Instituto de F\'isica, Universidade Federal do Rio Grande do Sul, 91501-970 Porto Alegre, Brazil \\
Journal: EPL \textbf{95}(5), 58006 (2011).
}

\abstract{
We present a general framework for the study of coevolution in dynamical systems.
This phenomenon consists of the coexistence of two dynamical processes on networks of interacting elements: node state change and rewiring of links between nodes. 
The process of rewiring is described in terms of two basic actions: disconnection and reconnection between nodes, both based on a mechanism of comparison of their states. 
We assume that the process of rewiring and node state change 
occur with probabilities $P_r$ and $P_c$ respectively, independent of each other. The collective behavior of a coevolutionary system can be characterized on the space of parameters $(P_r,P_c)$. As an application, for a voterlike node dynamics we find that reconnections between nodes with similar states lead to network fragmentation.
The critical boundaries for the onset of fragmentation in networks with different 
properties are calculated on this space. 
We show that coevolution models correspond to curves on this space describing functional relations between $P_r$ and $P_c$. The occurrence of a one-large-domain phase and a fragmented phase in the network
is predicted for diverse models, and agreement is found with some earlier results.
The collective behavior of system is also characterized on the space of parameters for the disconnection and reconnection actions. In a region of this space, we find a behavior where different node states can coexist for very long times on one large, connected network.}
\begin{document}
\maketitle

Many complex systems observed in nature can be described as dynamical networks of interacting elements or nodes where the connections and the states of the elements evolve simultaneously \cite{Zimmermann,Rohlf,Maxi,GrossR,GrossL}.
The links representing the interactions between nodes can change their strengths
or appear and disappear as the system evolves on various timescales. In many cases, these
modifications in the topology of the network occur as a feedback effect of the dynamics of the
states of the nodes: the network changes in response to
the evolution of those states which in turn determines the modification of the network. Systems that exhibit this coupling between the topology and states have been denominated as coevolutionary dynamical systems or adaptive networks \cite{Zimmermann,Maxi,GrossR}.

Coevolution dynamics has been studied in the context of spatiotemporal dynamical systems,
such as neural networks \cite{Kan1,Meisel}, coupled map lattices \cite{Kan2,Jap}, motile elements \cite{Kan3}, 
synchronization in networks \cite{Arenas}, 
as well as in game theory \cite{Zimmermann,Maxi,Gao}, spin dynamics \cite{mandra}, epidemic propagation \cite{gross,risau,Vasquez,shaw}, 
and models of social dynamics and opinion formation \cite{holme,Vasquez3,Centola,Vasquez2,kozma,kimura,medo}.

In many systems where this type of coevolution
dynamics is implemented, a transition is often observed from a phase
where most nodes are in the same state forming a large connected network to
a phase where the network is fragmented into small disconnected components, each
composed by nodes in a common state \cite{GrossF}. This network fragmentation transition
is related to the difference in time scales of the processes that govern the two dynamics: 
the state of the nodes and the network of interactions \cite{Vasquez2}.
In these models, the time scales of the processes of interaction between nodes 
and modification of their links are coupled and controlled by a single parameter in the system.

The phenomenon of coevolution raises one of the fundamental
questions in dynamical networks, namely whether the dynamics 
of the nodes controls the topology of the network, or this topology
controls the dynamics of the nodes.
In this paper we propose a general framework to approach this question. 
We consider that the process by which a node changes
its neighbors, called rewiring, and that 
the process by which a node changes its state,
have their own dynamics. Furthermore, we assume that these two processes can be independent of each other.
As a consequence of this assumption,
the collective behavior of a coevolutionary system can be studied on the space of the parameters representing the time scales for both processes. 
A particular coevolution dynamics can be described by formulating a specific coupling condition between the 
two competing processes in the network. 
We shall show that the collective behavior and the existence of a network fragmentation transition for given coevolution models can be predicted from the general phase diagram of the system on this space of parameters. 

Let us focus on the mechanisms for the rewiring process of the coevolution phenomenon.
For simplicity, we consider that the number of connections in the network is conserved. Then, we assume that any rewiring process consists of two basic actions: disconnection and reconnection between nodes. 
Both connecting and disconnecting interactions are often found in social relations, biological systems, and economic dynamics \cite{GrossR,GrossL,holme,kimura}. 

In general, either action, disconnection or reconnection, is driven by some mechanism of comparison of the states
of the nodes. We define a parameter $d\in [0,1]$ that measures the tendency to disconnect between nodes in identical states;
i.e., $d$ represents the probability that two nodes in identical states become disconnected and $1-d$ is the probability that two nodes in different states disconnect from each other. Similarly, we define another parameter $r\in [0,1]$ that describes the probability to connect between nodes in identical states; then, $1-r$ is the probability that two nodes in different states connect to each other.  A rewiring process can be characterized by the label $dr$, 
where $d$ indicates the probability for the disconnection action between nodes sharing the same state, and $r$ assigns the probability for reconnection between nodes possesing the same state.
Thus, we can construct a plane $(d, r)$ where any rewiring process subject to 
disconnection-reconnection actions between nodes can be represented as a point on this plane.

\begin{figure}[h]
\begin{center}
\includegraphics[scale=0.7,angle=0]{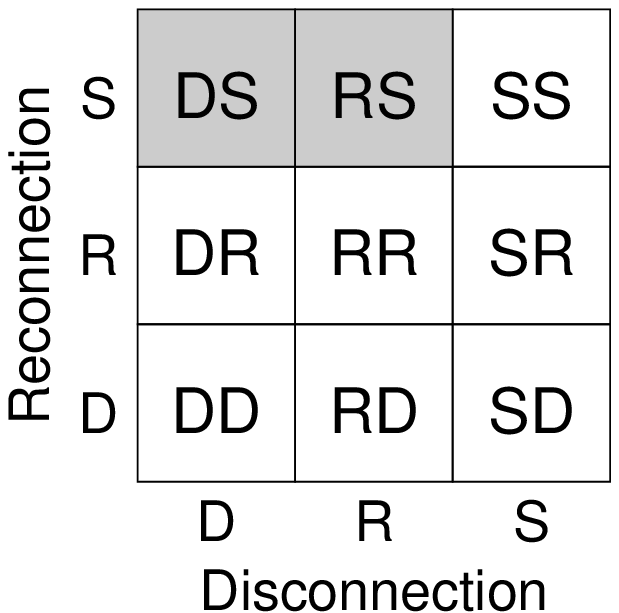}
\end{center}
\caption{Discrete rewiring processes on the disconnection-reconnection action space $(d, r)$. Either action can occur via three mechanisms: similarity (S), randomness (R), or dissimilarity (D). The two-letter labels describe the resulting rewiring processes $dr$. Rewirings that lead to a fragmentation transition in our model are colored in grey.}
\label{f1}
\end{figure}

In a simplified approach, we first consider a discrete expression of the plane $(d, r)$ as follows.
We assume that either action of the rewiring, disconnection or reconnection, can be driven by three distinct mechanisms: similarity $S$ (interaction between nodes sharing the same state), randomness $R$ (interaction between nodes regardless of their states), and dissimilarity $D$ (interaction between nodes having different states). Then both $r$ and $d$ can only take the values $0 (D)$, $0.5 (R)$, and $1 (S)$. This gives rise to nine possible rewiring processes based on the combinations of these actions and their mechanisms, as shown in Fig.~\ref{f1}. For example, $dr=RS$ denotes a rewiring where node $i$ is disconnected from node $j$ chosen at random and then reconnected to a node $m$ that possesses a state equal to that of $i$.  
We can classify many rewiring process employed in the literature under this scheme. 
For example, an $RS$ process corresponds to that used in Ref.~\cite{holme}, a $DS$ process was used in Ref.~\cite{Vasquez3}, while the rewirings employed in Refs.~\cite{Centola,Vasquez2,kozma} can be regarded as of type $DR$. Note that only the $RR$ process is completely independent of the states of the nodes.

Then a coevolutionary system can be analyzed as follows.
We assume that the dynamics of the system
can be described by the coexistence of a rewiring process $dr$
that takes place with a probability $P_r$, and a process of node state change that occurs with a probability $P_c$. 
We assume these two probabilities are independent of each other. 
Therefore, the dynamics of the coevolutionary system is represented by four basic parameters, $d, r, P_r, P_c$. 
The collective behavior of the system can be characterized on the space of these parameters. 
Then, a specific coevolution model associated to a rewiring process $dr$ consists of a 
prescribed functional relationship between the 
probabilities $P_r$ and $P_c$ that corresponds to a curve on the plane $(P_r,P_c)$.   

As an application of this scheme, consider a random network of $N$ nodes having average degree of edges $\bar{k}$, i.e., $\bar{k}$ is the average number of neighbors of a node. Let $\nu_i$ be the set of neighbors of node $i$, possessing $k_i$ elements. Let us assume that the network topology is subject to a rewiring process $dr$. 
For the node state dynamics, we choose a simple imitation rule such as a voterlike model that has been used in various contexts \cite{holme,Clifford,Holley,Castellano,Kra}.  
The state  of node $i$ is denoted by $g_i$, where $g_i$ 
can take any of $G$ possible options. The states $g_i$ are initially assigned at random with a uniform distribution.

The coevolution dynamics in this system is defined by iterating the following steps:
\begin{enumerate}
\item Choose randomly a node $i$ such that $k_i>0$.
\item With probability $P_r$, apply rewiring process $dr$: 
break the edge between $i$ and a neighbor $j \in \nu_i$ that satisfies mechanism $d$,
and set a new connection between node $i$ and a node $l \notin \nu_i$ that satisfies mechanism $r$.
\item Choose randomly a node $m \in \nu_i$ such that $g_i\neq g_m$.  With probability $P_c$, 
set $g_i=g_m$.
\end{enumerate}

Step 2 describes the rewiring process that allows 
the acquisition of new connections, 
while step 3 specifies the process of node state change; in this case
the states of the nodes becoming similar
as a result of connections. 
We have verified that the collective behavior of this system is statistically invariant if steps 2 and 3 are interchanged.
 
The network size $N$, the average degree $\bar{k}$, and the number of options $G$ remain constant during the evolution of the system.  Thus, given a rewiring process $dr$, the parameters of our model are the probability of rewiring, $P_r$, and the probability of changing the state of a node, $P_c$. 

The chosen imitation dynamics of the nodes tends to increase the number of connected pairs of nodes with equal states, while some rewiring processes may favor the fragmentation of the network. Therefore, the 
time evolution of the system should eventually lead to the formation of 
a set of separate components, or subgraphs,
disconnected from each other, with all members of a subgraph sharing the same state. 
We call {\it domains} such subgraphs.

\begin{figure}[h]
\begin{center}
\includegraphics[scale=0.31,angle=0]{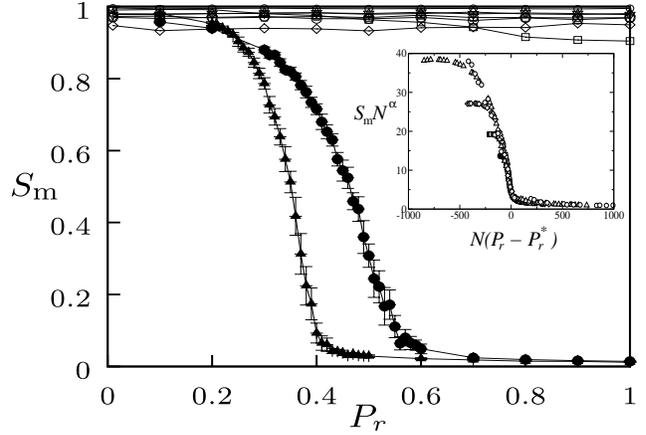}
\end{center}
\caption{$S_{\mbox{\scriptsize m}}$ as a function of $P_r$ for the $9$ rewiring processes in Fig.~\ref{f1}, with fixed $P_c=0.6$
and $G=320$. Network parameters are $N=3200$ and $\bar{k}=4$. Only rewiring processes $DS$ (triangles) and $RS$ (solid circles) exhibit a fragmentation transition.   
Error bars indicate standard deviations obtained over $100$ realizations of initial conditions
for each value of $P_r$. Inset: Scaling collapse found with the exponent $\alpha=0.5$, for the rewiring process $RS$ with
$P_c=0.6$. Sizes $N$ are $3200$ (circles), $1800$ (triangles), $800$ (diamonds), $400$ (squares), $200$ (solid circles}
\label{f2}
\end{figure}

To characterize the collective behavior of the system, we employ, as an order parameter, the normalized average size of the largest domain in the system, $S_{\mbox{\scriptsize m}}$.
Figure~\ref{f2} shows $S_{\mbox{\scriptsize m}}$ as a function of the probability $P_r$ for the discrete rewiring processes in Fig.~1 on a network having $\bar{k}=4$, with a fixed value of the probability $P_c$. 

We observe that most discrete rewiring processes in Fig.~\ref{f1} lead to collective states characterized by values $S_{\mbox{\scriptsize m}} \rightarrow 1$ and corresponding to a large domain whose size is comparable to the system size. 
However, the rewiring processes $DS$ and $RS$ exhibit a transition at some
critical value of $P_r$, from a regime having
a large domain, to a state consisting of only small domains for which 
$S_{\mbox{\scriptsize m}} \rightarrow 0$.  
Those rewirings $dr$ with $r=S$ can sustain a stable regime consisting
of many small domains ($SS$ leaves the initial network structure statistically invariant).
The critical point $P^*_r$ for the domain fragmentation transition in each case is estimated by
the value of $P_r$ for which the largest fluctuation of the order parameter $S_{\mbox{\scriptsize m}}$ occurs. 
For the rewiring process $RS$ on a network with $\bar{k}=4$, a finite size scaling analysis is shown in the inset 
in Fig.~2, where  $N^\alpha S_{\mbox{\scriptsize m}}$ is plotted versus $N(P_r-P^*_r)$, with $P^*_r=0.541 \pm 0.007$, and for various system sizes. We find that the data collapses in the critical region when $\alpha=0.50\pm 0.05$. 
A similar scaling analysis for the rewiring $DS$ in Fig.~2 yields $P^*_r=0.380  \pm 0.007$ and $\alpha=0.20 \pm 0.05$.
Thus, there exists a universal 
scaling function $F$ such that $S_{\mbox{\scriptsize m}}=N^{-\alpha}F(N(P_r-P^*_r))$ associated to each process $RS$ and $DS$.

\begin{figure}[h]
\begin{center}
\includegraphics[scale=0.43,angle=0]{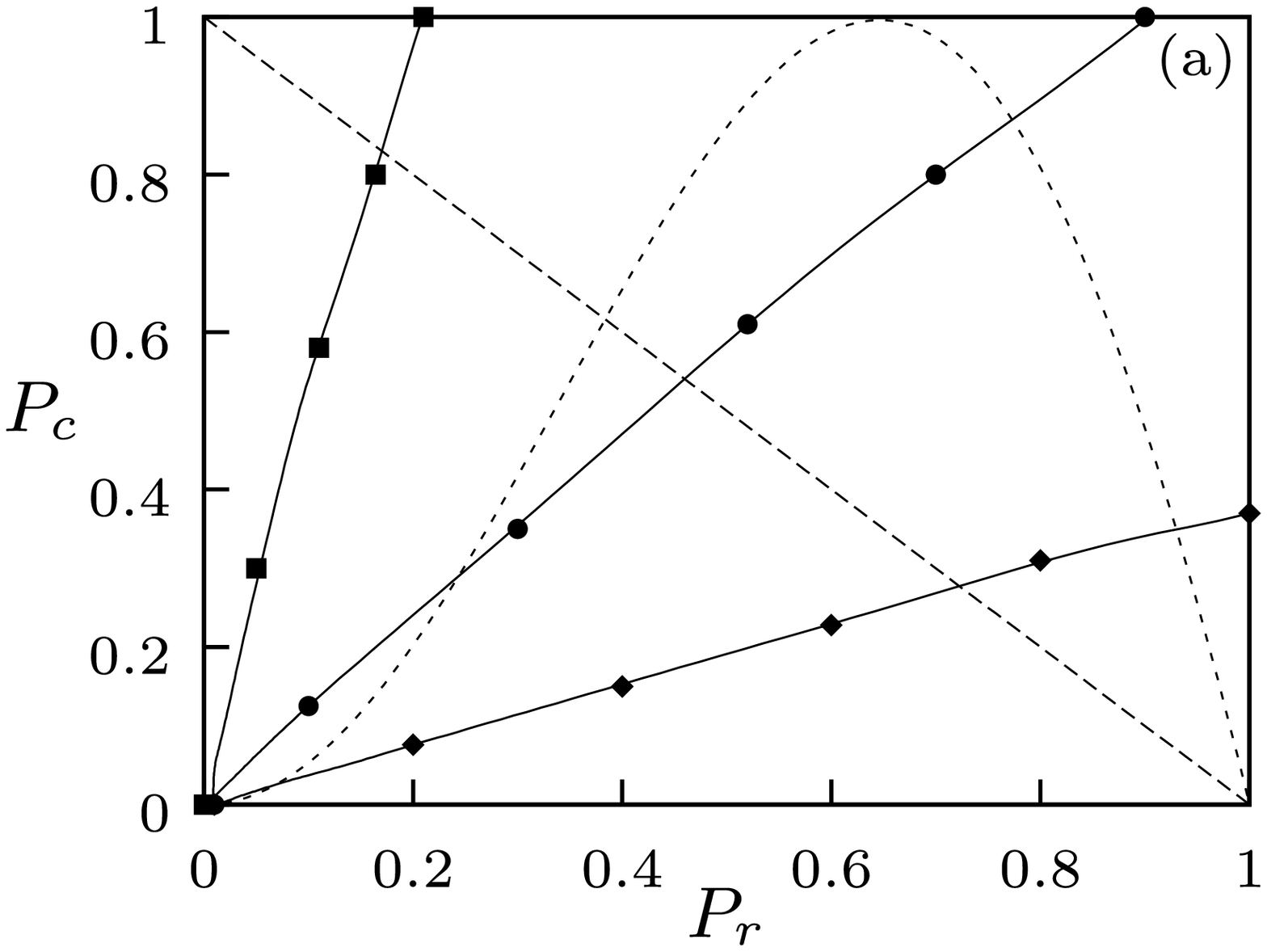} 
\end{center}
\begin{center}
\includegraphics[scale=0.43,angle=0]{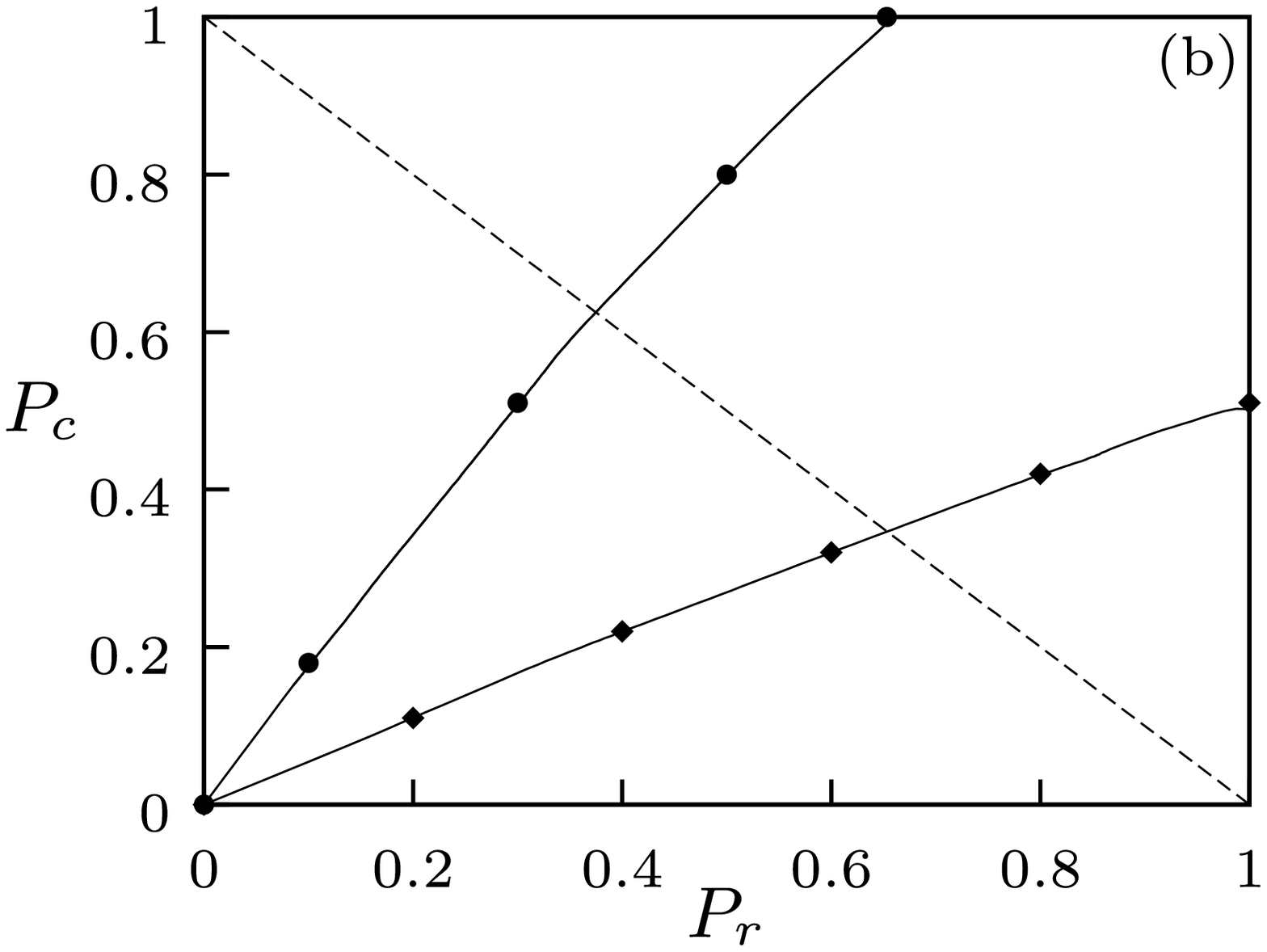}
\end{center}
\caption{Critical boundaries on the space of parameters $(P_r,P_c)$ for fragmentation transitions associated to two rewiring processes on a network of size $N=3200$. Each symbol-marked curve indicates the corresponding  boundary that separates the regions where a state having a large domain 
(above the curve) and a state consisting of many small domains (below the curve) occur. (a) Rewiring process $RS$ and node states with $G=320$ on a network having $\bar{k}=2$ (line with squares); $\bar{k}=4$ (circles); $\bar{k}=8$ (diamonds).
The slashed
line is the relation $P_c=1-P_r$, and the dotted line is $P_c=1.72 P_r\sin(\pi P_r)$.
(b) Rewiring process $DS$ and node states with $G=2$ on a network with $\bar{k}=4$ (line with circles); $\bar{k}=8$ (diamonds). The slashed
line is the function $P_c=1-P_r$.
All the numerical data points are averaged over $100$ realizations of initial conditions.}
\label{f3}
\end{figure}

For a given rewiring process, the collective behavior of the coevolving system can be characterized in terms of the quantity $S_{\mbox{\scriptsize m}}$ on the space of parameters $(P_r,P_c)$. Figures~\ref{f3}(a) and \ref{f3}(b) show the phase diagrams arising on the plane $(P_r,P_c)$ when the rewiring processes $RS$ and $DS$, respectively, are employed on networks having different values of $\bar{k}$. In both cases, for each value of $\bar{k}$, two phases appear in the system as the parameters $P_c$ and $P_r$ are varied: one phase consists of the presence of only small domains and is characterized by $S_{\mbox{\scriptsize m}} \to 0$, and the other is distinguished by the formation of a large domain and is characterized by larger values of $S_{\mbox{\scriptsize m}}$. These two regimes are separated by a critical curve $(P^*_c,P^*_r)$. 

Figure~\ref{f3} expresses the general phase diagram of a coevolving system subject to a given node state dynamics and a given rewiring process. Diverse coevolution models can be represented in this diagram by formulating specific coupling relations between the rewiring and the node state dynamics. In general, such a coupling can be expressed as a functional relation $P_c(P_r)$ that describes a curve on the space of parameters in Fig.~\ref{f3}. For example, consider the relation $P_c=1-P_r$
on the phase diagram in Fig.~\ref{f3}(a). 
This corresponds to the coevolution model proposed in Ref.~\cite{holme} that uses a rewiring of type $RS$. In this case, the transition from a large domain regime to a fragmented phase on a network characterized by a value of $\bar{k}$ should occur when this straight line intersects the corresponding critical boundary curve in Fig.~\ref{f3}(a). These intersections yield the values $P_r^*=0.171$ for $\bar{k}=2$, $P_r^*=0.458$ for $\bar{k}=4$, and $P_r^*=0.722$ for $\bar{k}=8$, which agree with the critical values found in \cite{holme}. Similarly, a rewiring of type $DS$ and the coupling function $P_c=1-P_r$ describe the two-state voter model introduced in Ref.~\cite{Vasquez3}. The intersection of the line $P_c=1-P_r$ with the boundary curve corresponding to $\bar{k}=4$ on the phase diagram in Fig.~\ref{f3}(b) indicates the critical value $P_r^*=0.375$. This value agrees with that calculated by a different procedure in Ref.~\cite{Vasquez3}. Furthermore, for a network having  $\bar{k}=8$, the predicted critical value for this model is $P_r^*=0.653$.

The phase diagrams of Fig.~\ref{f3} predict the critical values $(P_r^*,P_c^*)$ for the network fragmentation transition in more complicated coevolution models. For example, consider the nonlinear relation $P_c=a P_r \sin(\pi P_r)$ 
on the space of parameters of Fig.~\ref{f3}(a). 
For $a=1.72$, this function crosses the critical boundary associated to $\bar{k}=4$ in Fig.~\ref{f3}(a) twice, at the values $P_r^*=0.25$, corresponding to a recombination of the network, and $P_r^*=0.77$, signaling a fragmentation transition. In the range of parameters $P_r \in (0.25,0.77)$, the function lies within the one-large domain region of the phase diagram. Thus, in a coevolution model described by this function on a network characterized by $\bar{k}=4$, a regime of one large domain should exist for this range of parameters. For $\bar{k}=2$, only a fragmented phase occurs for this coevolution function.  

\begin{figure}[h]
\begin{center}
\includegraphics[scale=0.5,angle=0]{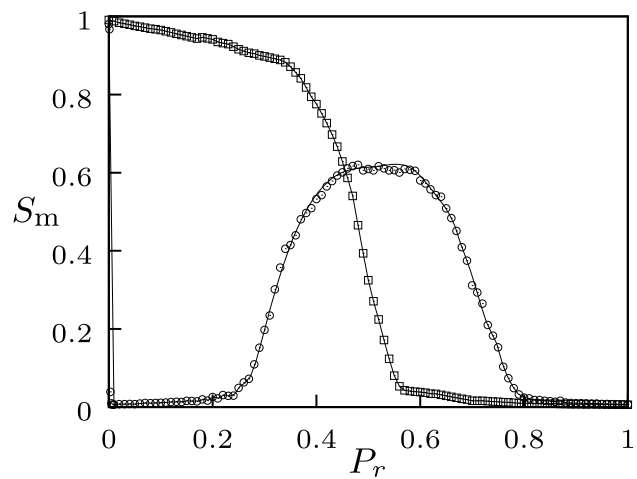}
\end{center}
\caption{$S_{\mbox{\scriptsize m}}$ as a function of $P_r$ for different coevolution curves subject to the rewiring process $RS$
in Fig.~3(a), on a network with $\bar{k}=4$.  $P_c=1-P_r$ (squares); $P_c=1.72P_r\sin(\pi P_r)$ (circles). For each value of $P_r$, $100$ realizations of initial conditions were performed.}
\label{f4}
\end{figure}

Figure~\ref{f4} shows $S_{\mbox{\scriptsize m}}$ as a function of $P_r$ for the two coevolution models
presented in Fig.~\ref{f3} for a network with $\bar{k}=4$. 
For the model in Ref.~\cite{holme}, the fragmentation transition takes place
at the value $P_r^*$ predicted from Fig.~\ref{f3}. 
Similarly, for the nonlinear model we confirm the existence of a one-large domain phase confined in 
the region $P_r \in (0.25,0.77)$. 

We have also investigated the behavior of the system on the space of parameters $(d,r)$ that describes general rewiring processes,
while keeping others parameters fixed. 
As before, we start from a random network and a random uniform distribution of states $g_i$. As an example, let us assume a dynamics such that $P_r=1$ (the rewiring process is always applied) and $P_c=1$ (nodes always copy the state of a neighbor). 
The above algorithm defining the coevolution dynamics can be employed as $d$ and $r$ are changed. 

\begin{figure}[h]
\begin{center}
\includegraphics[scale=0.3,angle=90]{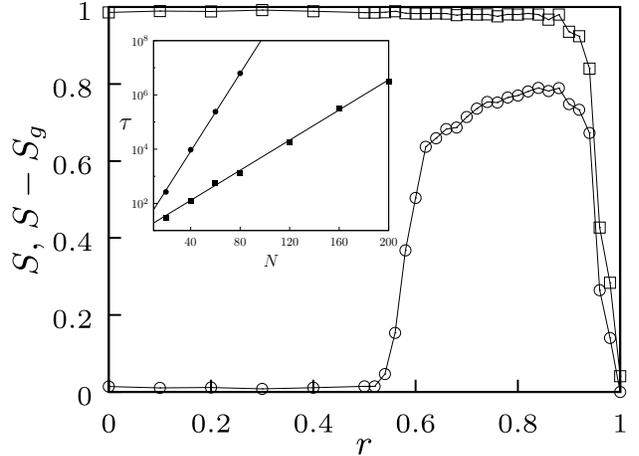}
\end{center}
\caption{$S$ (squares) and $S-S_g$ (circles) as functions of $r$, for rewiring process with fixed $d=0.2$, and $P_r=1$, $P_c=1$, $G=20$, $N=200$, and $\bar k=4$. For each value of $r$, $10$ realizations of initial conditions were performed.  
Inset: Semi-log plot of $\tau$ vs. $N$ for $r=0.2$ (solid squares) and $r=0.8$ (solid circles), with fixed $d=0.2$.
Each time step corresponds to $N$ iterations of the dynamics.}
\label{f5}
\end{figure}

Figure~\ref{f5} shows the average normalized size of the largest network component $S$, regardless of the states of the nodes, as a function of $r$, with fixed $d=0.2$. The quantity $S$ reveals a network fragmentation transition at a value $r=0.938$. We also calculate, for long times, the normalized average size of the largest subset of connected nodes in the largest network component that share the same state, denoted by $S_g$. Figure~\ref{f5} shows $S-S_g$ versus $r$. We observe that 
$S-S_g = 0$ for $r<0.56$, meaning that all the nodes on the largest component share the same state, on the average. Since $S\rightarrow 1$ for $r<0.56$, there is one large domain whose size is comparable to that of the system. For $r>0.938$, 
we have $S-S_g \rightarrow 0$ and $S\rightarrow 0$, corresponding to the occurrence of multiple small domains in the system. 
In the range $0.56<r<0.96$, we observe $S-S_g >0$, indicating that not all the nodes on the largest network component share the same state. Since $S\rightarrow 1$ in this range of $r$, the system there consists of a connected network whose size is comparable to the system size. Thus, in the range $0.56<r<0.938$ we find a situation where subsets having distinct states coexist
on a large connected network.  In order to elucidate the nature of this behavior, we show in the inset in Fig.~\ref{f5} a semilog plot of the average time $\tau$ for reaching one large domain ($S=S_g=1$) in the system versus the system size $N$, for different values of $r$. We find that $\tau$ scales exponentially with $N$ as $\tau \sim e^{\alpha N}$. Thus, the one-large domain phase cannot take place in an infinite size system. For a finite size system, the one-large multi-state component should eventually decay to one-large domain. We obtain numerically the exponents $\alpha= 0.064$ for $r=0.2$, in the one-large domain region, and $\alpha=0.167$ for $r=0.8$ in the one-large multi-state component region of Fig.~\ref{f5}. This means that the average decay time for the one-large multi-state component is several orders of magnitude larger than the corresponding time for the one-large domain phase. For $N=200$, our results imply convergence times of the order of $\tau\approx 10^6$ for $r=0.2$ and $\tau \approx 10^{14}$ for $r=0.8$. As $N$ increases, the decay of the one-large multi-state component cannot be observed in practice. Thus, our results
for continuous values of the parameters $r$ and $d$ of the rewiring process suggest a mechanism for the coexistence
of subsets of nodes having different states on a large connected network.

\begin{figure}[h]
\begin{center}
\includegraphics[scale=0.5,angle=0]{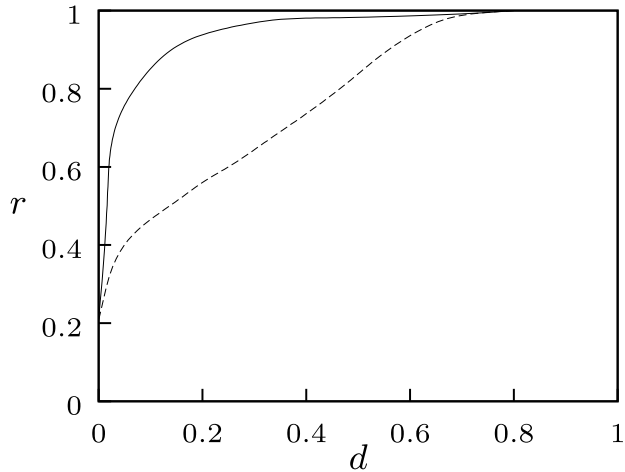}
\end{center}
\caption{Phase diagram on the space of parameters $(d,r)$, for $P_r=1$, $P_c=1$. Fixed $G=20$, $N=200$, and $\bar k=4$. The fragmented phase occurs above the continuous line; the one large-domain phase takes place below the dashed line; the region where one large-component with coexisting states emerges is bounded by these two lines. All numerical data points are averaged over $10$ realizations of initial conditions.}
\label{f6}
\end{figure}

For given values of $P_r$ and $P_c$ that describe a coevolution dynamics, the collective behavior of the system can be characterized on the space of parameters for the disconnection and reconnection actions, $(d,r)$,  by using the quantities calculated in Fig.~(\ref{f5}). Figure~\ref{f6} shows the phase diagram resulting on the plane $(d,r)$ for the values $P_r=1$ and $P_c=1$. Three types of behaviors occur in the system as the parameters $r$ and $d$ are changed. Two of these behaviors correspond to the phases already found in Fig~\ref{f3}: a one large-domain phase and a fragmented phase consisting of small domains. These two phases are separated by a region in the plane $(d,r)$ where a supertransient behavior emerges, characterized by the coexistence of several states on one large network component.
Figure~\ref{f6} reveals that the rewiring processes $RS$ ($d=0.5, r=1$) and $DS$ ($d=0, r=1$) yield a fragmented phase when $P_r=1$ and $P_c=1$, in agreement with the results found in Fig~\ref{f3}.

In conclusion, we have presented a general framework for the study of the phenomenon of coevolution in dynamical networks. Coevolution consists of the coexistence of two processes, node state change and rewiring of links between nodes, that can occur with independent probabilities $P_r$ and $P_c$, respectively. We have analyzed the process of rewiring 
in terms of the actions of disconnection
and reconnection between nodes, both based on a mechanism of comparison of their states.

For a given rewiring process, the collective behavior of a coevolving system can be represented in the space of parameters $(P_r, P_c)$.
For a voterlike node dynamics, we found that only reconnections between nodes with similar states can lead to network fragmentation. We have calculated the critical boundaries on this space for the fragmentation transition in networks having different values of $\bar k$.
The size of the region for the fragmented phase in the space $(P_r, P_c)$ decreases with increasing $\bar k$. This suggests that fragmentation is more likely to be observed in networks where $\bar k \ll N$. 
We have shown that coevolution models correspond to curves $P_c(P_r)$ on the plane $(P_r, P_c)$. The occurrence of network fragmentation as well as recombination transitions for diverse models can be predicted in this framework. 

We have also characterized the collective properties of the system on the space of actions for rewiring processes $(d,r)$, for given values of $P_r$ and $P_c$ that define a coevolution dynamics. On a region of this space, we have unveiled a regime where subsets having different states can coexist for very long times in one large, connected network .

We have limited our study to the case when then number of connections in the coevolving network is conserved. This condition is expressed in step 2 of the algorithm, where the application of both actions of disconnection and reconnection occurs with probability equal to one.  
This condition can be generalized by considering different probabilities for each of these actions. Thus, our framework provides a scenario for studying coevolving dynamical networks with no conservation of the total number of links. 

Other extensions to be investigated in the future include the characterization of the topological properties of the network on the continuous plane $(d, r)$, the consequences of preferential attachment rules for the reconnection action, the consideration of variable connection strengths,
and the influence of the node dynamics on the collective behavior of coevolving systems.

\acknowledgments
This work was supported by project C-1692-10-05-B from CDCHTA, 
Universidad de Los Andes, Venezuela. J.C.G-A was supported by project FISICOS FIS2007-60327 from M.E.C., Spain.
He also thanks CNPq, Brazil.

\end{document}